\pdfoutput=1
\documentclass{article}

\usepackage{natbib}
\usepackage[dblblindworkshop, final]{neurips_2026}

\usepackage[utf8]{inputenc} %
\usepackage[T1]{fontenc}    %
\usepackage{hyperref}       %
\usepackage{url}            %
\usepackage{xurl}           %
\usepackage{booktabs}       %
\usepackage{amsfonts}       %
\usepackage{nicefrac}       %
\usepackage{microtype}      %
\usepackage{xcolor}         %
\definecolor{linkblue}{RGB}{0,51,153}
\hypersetup{colorlinks=true, allcolors=linkblue}
\usepackage{cite}

\title{Protecting the System, Stranding the User: The Limits of Multi-Agent Authorization}

\author{%
  Gurpriya Bhatia \\
  University of Southern California \\
  Los Angeles, CA \\
  \texttt{gbhatia@usc.edu} \\
  \And
  Clifford Neuman \\
  University of Southern California \\
  Los Angeles, CA \\
  \texttt{bcn@isi.edu} \\
}
\makeatletter
\renewcommand{\@trackname}{Accepted as a poster at HAIPS: 2nd Workshop on Human-Centered Privacy and Security for Language Models, held in conjunction with COLM 2026.}
\makeatother

\begin{document}

\maketitle

\begin{abstract}
Current multi-agent authorization architectures prioritize system integrity over human outcomes. While recent models successfully prevent privilege escalation, they externalize the cost of failure onto the human operator. Re-centering authorization around human impact exposes four structural vulnerabilities in current paradigms. First, calling an agent's understanding of user intent verified is misleading. Intent has no formal definition, so no check against it can be definitive. Systems should instead report an openly probabilistic estimate and scale their caution with the risk of the action. Second, we argue that dynamic trust scores do not survive multi-agent handoffs; transmitting algorithmic judgment across a company boundary without underlying evidence risks violating Contextual Integrity. Third, treating a safely halted workflow as a successful refusal is a category error if it strands the user; compensating actions must be bound directly to the capability token, not only to the planning layer. Finally, we hypothesize that frictionless automated recovery introduces a moral hazard, disincentivizing developers from building robust initial intent checks. We conclude by proposing an architectural synthesis to bridge the gap between deterministic security and human outcomes, alongside the fundamental theoretical limits that remain unresolved.

\end{abstract}

\section{Introduction}
Consider a routine multi-agent delegation: a user instructs their personal assistant agent to secure a rental apartment. The agent is granted two explicitly scoped permissions: (1) pay a holding deposit up to a set limit, which is forfeited if the application does not complete, and (2) transmit only the required application documents. The ordering is deliberate and reflects standard letting practice: the holding deposit is what takes the property off the market, and referencing (identity, employment, and financial checks) follows once the unit is reserved. The assistant successfully executes the payment, then delegates the document collection task to the letting agency's intake agent. At this boundary, the chain breaks. A stale permission or a malicious prompt causes the agency's agent to overreach, demanding the user's complete bank history and biometric ID. A robust, out-of-model security monitor detects the violation and correctly terminates the workflow. The event is logged, and the system halts.

By every modern agentic authorization standard, this is a security success. The system evaluated the permissions, caught a scope violation, and safely aborted the action. Yet, for the human operator, the outcome is costly: the deposit is forfeited, the application is incomplete, and they are left holding a generic error message. One might object that the holding deposit is forfeited whether a human or an agent manages the transaction. But a human facing an unreasonable data request can negotiate, escalate, or choose to absorb the privacy risk to save their sunk cost. The agent security monitor cannot: it unilaterally aborts, never offering the user the chance to weigh the trade-off. The system protects the user's data by guaranteeing their financial loss.

This systemic rigidity exposes a deep blind spot in current privacy and authorization frameworks. The triggering overreach is, in fact, a Contextual Integrity violation \citep{barth2006privacy,nissenbaum2010privacy}: the agency's agent demands biometric identification and a full bank history, information whose flow is deeply inappropriate to the context of renting an apartment. Viewed through this lens, the user's security monitor did exactly what current theory prescribes: it refused a transmission that violated contextual norms. The deeper structural friction is that the two agents inhabit different contexts to begin with: a personal assistant and a commercial letting agency, and the norms of one do not govern the other. While Contextual Integrity provides the exact theoretical machinery to identify and refuse such an inappropriate flow, it says nothing about what the architecture owes the user once it has aborted. That silence, not the refusal, is where this paper begins.

This disparity exposes the central thesis of this paper: the field has designed authorization architectures around the integrity of the decision, while ignoring the outcome for the human. We argue that current multi-agent security paradigms successfully protect the system's state, but quietly externalize the cost of failure onto the user. To address the gap between system security and human outcomes, we advance four interconnected positions:
\begin{enumerate}
    \item \textbf{The illusion of verified intent:} Labeling an agent's semantic alignment as verified is a category error. Intent is not a formally defined property, so there is no definitive ground truth to confirm a match against; only noisy, post-hoc proxies such as user feedback, which measure tolerance rather than fidelity. The result is a guarantee the evidence cannot support, which obscures the uncalibrated risk thresholds underneath it.
    \item \textbf{The boundary problem of dynamic trust:} We argue that the probabilistic trust scores currently proposed to solve intent alignment do not survive a cross-company handoff without their underlying evidence. Passing a bare trust score across a context boundary is worse than passing nothing if naively trusted.
    \item \textbf{Token-bound recovery:} We argue that correctly halting a broken chain is a category error if it strands the user. Recovery should also push the Sagas pattern down into the authorization layer, embedding the compensating action directly inside the capability token.
    \item \textbf{The paradox of safe recovery:} Drawing on risk compensation in safety engineering, we warn that if these token-bound recoveries become too cheap and reliable, they may introduce a moral hazard that incentivizes developers to weaken initial intent checks.
\end{enumerate}

While recent work has made vital strides in formalizing authorization propagation \citep{tallam2026authorization}, these models stop at the system boundary. They dictate how to stop the damage, but not what the system owes the user once the deposit is spent and the workflow fails.

Section 2 outlines the settled baseline of agentic security on which we build. Section 3 details our four core positions regarding the human-system boundary. Section 4 proposes an architectural synthesis to bridge these gaps, and Section 5 transparently addresses the theoretical limits we cannot yet resolve.

\section{Prior Work: What Is Settled and What Is Emerging}

Several subfields are converging on agent authorization. We separate two types of prior work. The first is genuinely settled: the reference monitor, the separation of decision from enforcement, the Sagas pattern, and classical trust management are decades old and well validated. The second is an emerging consensus: agentic systems and standards from the past year that share a common architecture but lack independent validation. Our critique targets this second group. Four of the subsections below pair a proven principle with its recent agentic application: the reference monitor with runtime kernels, verification limits with dynamic risk scoring, capability attenuation with delegation protocols, and Sagas with agent recovery; a fifth surveys industry standardization.

\paragraph{Separation of Decision and Enforcement:}
There is broad consensus that a model should not authorize its own actions. Probabilistic compliance is insufficient for security; the final authorization must occur outside the model via deterministic code \citep{bhattarai2026deterministic}. This mirrors the 1970s reference monitor concept, which isolates a trusted component to mediate all access \citep{anderson1972computer, saltzer1975protection}. Recent systems like KAIJU \citep{guerin2026kaiju} enforce this at runtime, ensuring the model never dictates its own intent. The industry AuthZEN \citep{openid2026authzen} standard similarly separates decision from enforcement. We take this architectural boundary as a settled baseline.

\paragraph{The Limits of Formal Verification:}

Even with external checks, defining what to check remains a fundamental challenge. Shi et al.~\citep{shi2025sok} systematically review agent security and, drawing on Rice's Theorem \citep{rice1953classes}, argue that verifying whether an agent's behavior satisfies a given non-trivial semantic property across all inputs is mathematically undecidable. Consequently, the field is pivoting toward dynamic risk assessment, weighing an agent's real-time trustworthiness against action risk. This pivot has deep roots in early trust-negotiation and trust-management frameworks. Role-based trust management (RT) \citep{li2002design} established how to formally reason about credentials and delegated authority across separate administrative domains, while ATNAC \citep{ryutov2005adaptive} introduced dynamic adjustment, scaling required credentials against a calculated suspicion level for each counterparty. These frameworks excel at evaluating deterministic, cryptographic credentials, but provide no machinery for open-ended semantic intent-fidelity. This represents the precise gap around unverifiable intent that Position 1 develops.

\paragraph{Delegation Infrastructure:}
The mechanisms for transferring authority between agents are maturing. Tallam \citep{tallam2026authorization} formalizes authorization propagation as an identity-governance problem, requiring access scopes to attenuate at each hop. SentinelAgent \citep{patil2026sentinelagent} formalizes delegation chains as a calculus of verifiable properties enforced by a runtime authority; notably, its own analysis converges with Shi et al.~\citep{shi2025sok}, concluding that deterministic intent verification is infeasible and treating intent as the single property it can check only probabilistically.

\paragraph{Industry Standardization:} Commercial deployment of these concepts is accelerating. In early 2026, Mastercard and Google released Verifiable Intent \citep{mastercard2026verifiable}, an open standard cryptographically tying agent purchases to human authorization. Concurrently, standards bodies have taken up agent identity directly: the FIDO Alliance has begun developing standards for trusted AI agent interactions through its technical working groups \citep{fido2026agentic}, while NIST's NCCoE has proposed a demonstration project applying identity standards to agent identification, delegation, authorization, and logging \citep{nccoe2026agent}.

\paragraph{Agentic Recovery Mechanisms:} Finally, distributed systems recovery patterns are being adapted for agents. SagaLLM \citep{chang2025sagallm} ports the Sagas transactional pattern into multi-agent LLM planning, pairing workflow steps with automated compensation steps. Checkpoint-restore mechanisms are also being applied to rewind agents, though naive re-execution after a restore is unsafe \citep{zheng2026acrfence}. However, as we will argue, this recovery is currently isolated at the planning and runtime layers, ignoring the underlying token layer.

The classical foundation is well established, and a consensus is forming quickly on top of it. But critical vulnerabilities remain in how these chains handle context, failure, and guarantees.

\section{Four Vulnerabilities in the Human-System Boundary}

These four positions are not independent complaints. Position 1 establishes what an authorization decision actually carries: not a verified fact about the user's intent, but an estimate. Positions 2, 3, and 4 follow that estimate as authority moves: across a context boundary, through a failure, and back into the next cycle of trust.

\subsection{Position 1: The Illusion of Verified Intent}
Authorization gates occur at a highly specific moment in an agent's workflow: after an action is proposed, but before it impacts the external world. In transactional terms, this is the commit point for the action, that is, the precise instant after which the operation cannot be trivially aborted, and any subsequent failure will require structural compensation. When evaluating the proposed action at this gate, the field currently conflates two fundamentally distinct questions, relying on the single term verified to describe both.

\textbf{The first question is quantitative:} Would the action follow the rules? Would it stay under budget, restrict itself to permitted files, and only call authorized tools? This is a deterministic bounds check with a clean yes or no answer. \textbf{The second question is semantic:} Would the action do what the user actually wanted? Shi et al.~\citep{shi2025sok} invoke Rice's Theorem to argue that verifying an agent's behavior against a well-defined semantic property is undecidable. Our claim does not rest on that result. Intent is not a formally defined property at all, so there is no ground truth against which a check could be definitive, only noisy proxies such as user feedback. Treating semantic intent as verifiable is therefore a category error: it can only ever be a probabilistic estimate.

Critically, these two questions suffer from a severe temporal asymmetry. Rule-conformance has a definitive after: a system can predict if a tool call will go over budget, and later confirm whether it actually did. Intent has no such after. A system can estimate intent before the action, but it can never definitively confirm it afterward because the necessary ground truth, that is the internal state of the human's mind, is unobservable. While systems collect user feedback or corrections post-execution, these are noisy, retrospective proxies heavily influenced by outcome bias, not ground truth.

Because intent lacks an after, the term verified is not just loose shorthand; it is misleading. Verified implies an after the fact confirmation that a claim is true. By applying it to semantic intent, the field conflates the certainty of the quantitative check with the fundamental ambiguity of the semantic one. The Mastercard and Google `Verifiable Intent' standard \citep{mastercard2026verifiable} perfectly illustrates this hazard. The protocol's underlying engineering is cryptographically sound: it binds the quantitative rules (amount, merchant, time window) and carries the open-ended user instructions without claiming cryptography enforces them. However, the nomenclature risks misleading auditors, regulators, and users by implying that the semantic alignment problem has been mathematically solved when it remains fundamentally open.

Our first position is that the field must decouple these concepts. Systems must report a clean pass/fail for rule-conformance, and a separately labeled, openly uncertain intent-fidelity score. An obvious objection is that if we cannot formally validate an intent score, why report it? Is this not just relabeling the problem? We argue that the value of the score lies in transparency. A system that explicitly states, ``We estimate this matches your intent, and here is our confidence level,'' places the uncertainty where the user can see it. The word verified hides that exact same uncertainty behind a false promise of safety. An explicitly probabilistic estimate is more rigorous than an unprovable guarantee.

\subsubsection{The Three Problems with Thresholds}

An intent-fidelity score is ultimately just a number, and eventually the system must use it to make a binary act-or-block decision. Shi et al.~\citep{shi2025sok} attempt this by combining the strength of the evidence with the reliability of its source and checking the result against a cutoff dictated by the action's risk.

Although structurally correct, the calibration of this cutoff remains unspecified, hiding three major vulnerabilities:

\begin{enumerate}
    \item \textbf{Unanchored Risk Scaling}: If the system's confidence threshold must scale with risk, we must establish a concrete metric for that risk. We argue that risk should be anchored directly to reversibility (detailed in Position 3). The structural or physical permanence of an action should dictate the minimum confidence score required to execute it. Prior systems scale credential requirements with transaction risk (e.g., ATNAC \citep {ryutov2005adaptive}), and Shi et al.~\citep {shi2025sok} include operational irreversibility as one of three factors in their risk metric. Our claim is narrower: reversibility should not be one input among several but the primary anchor, because it is the one property that determines whether the failure modes in Position 3 have any remedy at all.

    \item \textbf{The Calibration Paradox}: Calibrating a threshold means checking it against outcomes. However, because intent lacks an observable after, system designers lack the ground truth required to objectively tune the cutoff. While developers routinely rely on post-hoc user feedback or workflow acceptance rates to tune these models, this telemetry measures user complacency and outcome tolerance, not the strict semantic fidelity of the original intent. We are attempting to calibrate a security threshold using a noisy proxy for the exact quantity this paper argues we do not formally possess.

    \item \textbf{Warning Fatigue and Human Failure}: The two ways of failing
    a threshold check carry asymmetric visibility. Setting the cutoff too low
    lets misaligned actions execute, the visible harm this paper addresses.
    Setting it too high means the system constantly stops to ask the user for
    confirmation. While this second failure looks safer on paper, usable-security research demonstrates that adherence to security warnings is far from reliable in practice, a pattern known as warning fatigue \citep{akhawe2013alice, anderson2015polymorphic}. A threshold tuned exclusively to avoid machine error quietly manufactures
    human error, training users to blindly rubber-stamp prompts.
\end{enumerate}

Therefore, our recommendation is twofold: first, report rule-conformance and intent-fidelity as strictly separate results. Second, treat the intent threshold not as a simple mathematical filter, but as a sociotechnical design decision that is tied to action reversibility, acknowledging the calibration paradox, and built to avoid both machine overreach and human warning fatigue.

\subsection{Position 2: The Boundary Problem of Dynamic Trust}

Evaluating dynamic trust instead of checking static permissions is designed for a single agent making a single decision. However, in practice, tasks are rarely confined to a single agent. An assistant agent routinely delegates sub-tasks to external agents like an intake agent run by a completely different letting agency. We must ask: what actually survives this handoff?

The quantitative permissions survive, and the infrastructure for this is well understood. Systems like the Human Delegation Provenance protocol (HDP) \citep{dalugoda2026hdp} cryptographically record that each hop in a delegation chain traces back to a human authorization event and its declared scope, and the structural requirements formalized by Tallam \citep{tallam2026authorization} call for access scopes to attenuate at each hop. But both preserve what was authorized, not how confident an upstream agent was that a specific action matches that authorization.

It is that dynamic judgment, the probabilistic intent-fidelity score established in Position 1, that does not survive the handoff. SentinelAgent \citep{patil2026sentinelagent} provides a robust framework for enforcing intent-preservation (the faithful downstream propagation of authorization parameters) through a single runtime authority with visibility across the chain. However, propagating authorization limits is distinct from transmitting an upstream agent's subjective confidence. SentinelAgent's approach is highly effective within a single trust domain. As multi-agent systems expand across company boundaries, however, shared visibility naturally ends. Addressing this cross-boundary handoff where dynamic judgment must travel without a shared authority represents the next frontier for this architecture.

An intent-fidelity score is a number, making it computationally cheap to attach to a token. However, everything that gives that number meaning stays behind: the conversation the first agent parsed, its internal weighting of evidence, and the reasoning that produced the score. The receiving agent has three options. It can blindly trust the number, inheriting a judgment made under foreign norms. It can rebuild the judgment from scratch, which is computationally expensive and often impossible because it lacks the original user input. Or, the sending agent can transmit the underlying evidence along with the score, which requires a massive privacy leak of one company's internal conversation logs into another's system. This echoes the classic trust-negotiation dilemma, where disclosing credentials to satisfy a stranger's policy itself risks leaking sensitive information \citep{winsborough2004safety}.

Because of this trilemma, passing a bare trust score across a company boundary is worse than passing nothing, if naively trusted. Passing nothing forces the receiving agent to realize it lacks context and must make its own judgment. Passing a naked score is dangerous because it looks like checkable data when it is not; unable to reconstruct it, engineers tend to treat the unverifiable estimate as objective fact. Accepting a score on blind faith quietly recreates the exact static-permission problem the field is trying to escape, merely relabeled as `dynamic trust'. A system that cannot independently assess a cross-boundary trust signal should refuse to act on it.

\paragraph{Contextual Integrity: Data Flow vs. Judgment Flow}
This is not merely an engineering inconvenience; it is a Contextual Integrity problem once that framework's logic is extended from data flows to judgment flows. Here, the company boundary is also a context boundary, though the two are not always the same, and it is the context boundary doing the work. The first agent formed its judgment under one specific set of norms. The receiving agent operates under entirely different norms, and there is no reason a trust score summarizing the former should hold true for the latter.

Recent literature has explored Contextual Integrity to govern whether a flow of data is appropriate across agentic boundaries \citep{bagdasarian2024airgapagent,ghalebikesabi2025privacy, mireshghallah2024can, shao2024privacylens}. Closest to our concern, Abdelnabi et al.~\citep{abdelnabi2025firewalls} introduce the use of dual firewalls between communicating agents, projecting messages onto the task's context to ensure only appropriate content crosses. This is a vital application of the theory. All of this work, however, governs whether a flow of data is appropriate; none of it addresses whether an upstream agent's probabilistic judgment about intent can securely survive a move between contexts. We argue that this protection can be strengthened by expanding its scope from content filtering to judgment transmission.

The boundary problem has a concrete engineering consequence. Caching a prior check's result is a natural optimization, but it only works intra-context, because the same party is asking the same question twice. It breaks the moment the asking party changes. Dynamic, risk-aware trust has so far been designed for a single decision point. Whether it carries any meaning once the judgment crosses into another agent, system, or company remains an unsolved problem. For cross-company handoffs, transferring dynamic trust without transferring the underlying data may be impossible in principle rather than merely hard. This is a conjecture, not a theorem, and one the literature has yet to formally examine.

\subsection{Position 3: Token-Bound Recovery vs. Correct Refusals}
When a distributed agentic chain breaks or a trust check fails, the field's default solution is fail-and-report: stop the action and log the error. Systems like  SentinelAgent \citep{patil2026sentinelagent} enforce this to keep damage within bounded limits. However, calling this a successful refusal mislabels the problem. Security engineers measure the decision; the human user measures the result. If an agent correctly stops a workflow mid-execution because a trust threshold was missed, the outcome for the user is still a failure. The holding deposit is forfeited, the application is incomplete, and all the user holds is an error message. A correct refusal that leaves the user stranded is a systemic failure, not a security success.

The fundamental technique for handling multi-step failures without atomic locks was solved decades ago: the Sagas pattern \citep{garcia1987sagas}. Sagas break a process into steps and pair each step with a pre-defined compensating action (an undo or remedy) that automatically executes if a later step fails. As noted earlier, work like SagaLLM \citep{chang2025sagallm} has recently brought this pattern to agent systems. Our position is not simply that the field should borrow Sagas; it already has. Our claim is more specific and highly contestable: compensating actions belong inside the capability token, not only in the workflow plan.

Current systems like SagaLLM define compensations at the planning layer, assuming the agent knows the workflow from end to end. As Position 2 argued, this assumption breaks upon delegation. The letting agency's agent was never part of the user's original plan; the only thing that travels across that boundary is the token. If the compensating action is not bound to the token, it does not cross the boundary, leaving the remote agent with no remedy to execute. This principle builds directly on insights from the runtime layer. ACRFence \citep{zheng2026acrfence} effectively demonstrates that naive re-execution is unsafe: because an agent re-synthesizes a subtly different request after a restore, a retried tool call can trigger semantic rollback attacks. Their work shows that recovery is much harder than simply retrying a prompt. While ACRFence successfully secures the tool boundary inside a single runtime, cross-agent workflows require this same level of protection to travel with the authorization itself. By embedding the remedy directly into the capability token, we complement planning-layer solutions like SagaLLM \citep{chang2025sagallm} and runtime-layer solutions like ACRFence, ensuring that authority crossing a delegation boundary is always accompanied by its compensating action. One might object that compensation is a workflow concern, not an authorization concern. We disagree. If a capability token carries the authority to incur a real-world cost, it must also carry the authority and instructions for the remedy.

Because the field has ignored the token layer for recovery, three critical infrastructure gaps remain:

\begin{enumerate}
    \item \textbf{Remedy-less Tokens}: None of the dominant capability-token
    formats, such as Macaroons \citep{birgisson2014macaroons}, records what to do if the
    permission is revoked post-execution. The remedy should be decided at mint
    time and travel with the authority, rather than being improvised by
    whichever remote agent happens to catch the error.

    \item \textbf{The Propagation Problem}: To trigger a compensation, a system
    must be able to tell every branch of a delegation tree that an action has
    been aborted. Bearer tokens (like the macaroon family) are valuable precisely
    because they can be verified offline without a central check. However, this
    offline nature makes immediate revocation or update across distributed nodes
    difficult without reintroducing an online check (for example, a third-party
    caveat), complicating multi-agent recovery efforts.

    \item \textbf{Forensic Logs vs. Human Decisions}: When an automated recovery
    fails, a human must intervene. Current systems generate detailed audit logs
    designed for investigators looking back months later. They do not generate
    actionable summaries for a user under time pressure. Furthermore, the
    literature tends to assume the original user is instantly available to
    receive the fail-and-report ping. In reality, the user may be asleep
    or offline, raising unsolved questions about asynchronous human oversight
    and escalation. The field has confused having a record with supporting
    a decision.
\end{enumerate}

\paragraph{The Limit of Reversibility}
There is a hard boundary to this approach: Sagas can only handle the reversible parts of a failure. A token can authorize the release of a held budget or the cancellation of a pending charge, but it cannot recover a forfeited holding deposit. This limitation is not a flaw in the Sagas pattern; it is a reality of interacting with the physical world. This is exactly the situation in our opening example: no compensating action can return the user's deposit. For irreversible steps like this one, what the architecture owes the user is not a rollback but a decision. Before aborting, the system should use the actionable summary proposed in Section 4 to surface the trade-off the monitor would otherwise resolve unilaterally, so that the user, not the monitor, chooses which loss to accept. And as we will argue in Position 4, when systems make the reversible parts of a workflow exceptionally cheap to undo, it fundamentally alters how agents and humans treat the irreversible parts.

\subsection{Position 4: The Paradox of Safe Recovery}
While distributed recovery patterns are maturing, the systemic relationship between transactional rollbacks and authorization state remains largely unexamined in the multi-agent literature. If a system can cheaply and reliably undo a mistake, does that change how hard it tries to avoid the mistake in the first place? In economics, this is known as moral hazard: shielding a party from a cost incentivizes them to take greater risks. In safety engineering, the same effect is called risk compensation, a long-debated phenomenon where introducing safety improvements (like seatbelts or better brakes) leads people to take more risks, offsetting the safety gains \citep{peltzman1975effects, wilde1982theory}. We argue that building robust agentic recovery mechanisms without accounting for this dynamic is a critical oversight.

Consider the interaction between the intent-fidelity thresholds discussed in Position 1 and the token-bound Sagas recovery proposed in Position 3. If a held budget is automatically released and a failed application is simply retried in the background, the immediate cost of a mistake drops to near zero. This creates a dangerous incentive for system designers. As noted in Position 1, asking a user to confirm every action causes warning fatigue. If recovery is cheap, developers may be tempted to lower the agent's intent-fidelity threshold to minimize user friction. The agent is permitted to guess, act, and fail more frequently because the Sagas pattern quietly mitigates the resulting errors. In this scenario, robust recovery acts as a systemic subsidy for careless intent alignment. The party that benefits from the shortcut (the developer optimizing for speed and low friction) is disconnected from the party that ultimately pays for the infrastructure, compute, and systemic risk of constant rollbacks. This is the classic condition for moral hazard. This is not an argument against the recovery mechanisms proposed in Position 3; they remain absolutely necessary. Rather, it is an argument that a recovery protocol requires a parallel feedback mechanism.

Even when an action's effects are fully and cleanly undone, the fact that a rollback occurred must leave a permanent mark that shapes future trust judgments. In aviation, safety programs meticulously record near-misses even when no one was harmed, precisely to ensure the next close call is less likely.
Agentic systems require a similar architecture.
A triggered compensation step must negatively impact the reputation score of the agent that initiated the failure: a standing, per-agent record, distinct from the per-action intent-fidelity score discussed in Positions 1 and 2. Earlier trust systems did track failures to adjust trust: ATNAC \citep{ryutov2005adaptive} raised a requester's suspicion score on bad-credential or probing events, tightening future policy. But that signal targets an adversarial counterparty; what remains absent in multi-agent authorization is a feedback channel that turns a recovery or rollback event into a lasting trust penalty against a cooperating upstream agent and the authority that issued its token.

\section{Architectural Synthesis: Toward Human-Centered Authorization}
None of the four positions presented here requires abandoning the consensus architecture the field has converged upon: cryptographically signed user intent, attenuating capability tokens, and deterministic, out-of-model reference monitors. Our critique is not that the current architecture is wrong, but that it successfully protects the system while externalizing the cost of failure onto the human user.

To re-center the architecture around human outcomes, we propose four structural additions:
\begin{enumerate}
    \item \textbf{Decoupled Reporting and Risk-Anchored Thresholds} (Position 1):
    Abandon the false guarantee of verified intent. Systems must report a
    binary pass/fail for rule-conformance, alongside an explicitly probabilistic
    intent-fidelity score. The threshold for action must not be static; it must
    be tied directly to action reversibility, balancing the
    risk of machine overreach against the human hazard of warning fatigue.

    \item \textbf{Evidence-Bound Handoffs} (Position 2):
    To preserve Contextual Integrity, a dynamic trust score must never cross
    a company boundary as a naked number. When authority crosses a context
    boundary, the protocol must enforce a strict binary: either transmit a
    cryptographic summary of the underlying evidence (without leaking the
    upstream agent's full internal logs, if such a summary proves achievable; see Section 5), or drop the score entirely and force
    the receiving agent to make its own judgment.

    \item \textbf{Token-Bound Sagas and Actionable Summaries} (Position 3):
    Recovery cannot live only at the workflow planning layer; it must also live in
    the authorization layer. Every permission that carries a real-world effect must
    be bound to a compensating action (the Sagas pattern) at the moment the
    capability token is minted, ensuring the remedy travels with the authority.
    Furthermore, when a chain terminates, or is about to halt with an irreversible loss at stake, the system must generate a localized,
    human-readable summary (what completed, what was undone, what is pending,
    and what is irreversible) distinct from forensic audit logs.

    \item \textbf{Recovery Penalty Feedback Loops} (Position 4):
    To counteract the moral hazard of cheap recovery, the authorization protocol
    must include a feedback channel. When an agent's token-bound compensating
    action is triggered, that failure must permanently register against the
    upstream agent's reputation score. System designers must not be subsidized
    for careless intent alignment by frictionless automated rollbacks.
\end{enumerate}

\section{Unresolved Theoretical Limits: What We Don't Know}
We conclude by transparently identifying the boundaries of our own proposals. The following challenges represent fundamental limits in agentic authorization that we do not yet know how to solve:

\begin{enumerate}
    \item \textbf{The Calibration Paradox:} As established in Position 1, intent lacks an observable after, so the ground truth needed to calibrate an intent-fidelity threshold may be permanently out of reach. We do not yet have the tools to close the gap between a transparent probabilistic estimate and a formally validated threshold.

    \item \textbf{The Cross-Boundary Compression Limit:} We do not know if a trust signal can be compressed enough to cross a context boundary without either leaking excessive proprietary data or shedding its semantic meaning. For cross-company handoffs, transferring trust without violating privacy may represent a fundamental cryptographic limit rather than a mere engineering challenge.
    \item \textbf{Physical and Market Irreversibility:} Proving that a token-bound compensating action restored the world is an open problem. A pending financial charge or a database write rolls back cleanly. An inappropriately read email, or a market price that competitors have already reacted to, cannot be undone by any protocol.
    \item \textbf{Distributed Legal Liability:} When a workflow spans multiple organizations, ownership of the token-bound recovery is unsettled. If a letting agency's agent triggers a failure, but the user's personal agent minted the compensation token, who legally owns the execution of the remedy? This is a liability and contract problem that technical capability tokens cannot currently express.
    \item \textbf{Asynchronous Escalation:} We do not know how a recovery summary should be intelligently routed and simplified when a delegation chain runs so deep that no single human could understand the full context under time pressure, especially if the original authorizing user is offline.
    \item \textbf{Measuring Risk Compensation:} Finally, the moral hazard proposed in Position 4 remains an architectural hypothesis. While work on risk compensation in safety engineering suggests that frictionless recovery could incentivize weaker intent checks, that effect remains debated even in its original domain, and we have not yet empirically measured it in production multi-agent systems. We present it here because the field is rapidly building the exact conditions necessary for it, and the effect has not been measured.

\end{enumerate}

\section{Conclusion}
The past year has seen rapid convergence on a formidable problem: isolating the model from the authorization decision, strictly attenuating delegated authority, and, in the research literature, accepting that human intent, unlike a budget limit, is not a property that can be deterministically verified at all. Yet, every one of these advances prioritizes the integrity of the system's internal state. None of them centers the human operator who bears the real-world cost of those decisions. That asymmetry is the gap this paper stakes out.

To bridge it, the field must answer four uncomfortable questions: What do we call a probabilistic semantic alignment, instead of falsely branding it as verified? What meaning does a dynamic trust judgment hold once it crosses a context boundary without its underlying evidence? What does an architecture owe a user after it correctly aborts a workflow mid-execution? And does frictionless, automated rollback introduce a moral hazard that quietly incentivizes developers to weaken their initial intent checks?

The first and the last are the questions the field has been most reluctant to confront, precisely because they are where the human operator is most exposed. Ultimately, a system that enforces the rules perfectly but abandons the user in the fallout has not solved agentic authorization. It has only solved the easy half.

\small
\bibliographystyle{plain} %
\bibliography{{custom}}

\end{document}